# A Pragmatic Methodology for Blind Hardware Trojan Insertion in Finalized Layouts

Alexander Hepp
alex.hepp@tum.de
Technical University of Munich
Department of Electrical and
Computer Engineering
Munich, Germany

Tiago Perez
Samuel Pagliarini
tiago.perez@taltech.ee
samuel.pagliarini@taltech.ee
Tallinn University of Technology
Department of Computer Systems
Tallinn, Estonia

Georg Sigl
sigl@tum.de
Technical University of Munich
Department of Electrical and
Computer Engineering
Munich, Germany
Fraunhofer AISEC
Munich, Germany

## ABSTRACT

A potential vulnerability for integrated circuits (ICs) is the insertion of hardware trojans (HTs) during manufacturing. Understanding the practicability of such an attack can lead to appropriate measures for mitigating it. In this paper, we demonstrate a pragmatic framework for analyzing HT susceptibility of finalized layouts. Our framework is representative of a fabrication-time attack, where the adversary is assumed to have access only to a layout representation of the circuit. The framework inserts trojans into tapeout-ready layouts utilizing an Engineering Change Order (ECO) flow. The attacked security nodes are blindly searched utilizing reverse-engineering techniques. For our experimental investigation, we utilized three crypto-cores (AES-128, SHA-256, and RSA) and a microcontroller (RISC-V) as targets. We explored 96 combinations of triggers, payloads and targets for our framework. Our findings demonstrate that even in high-density designs, the covert insertion of sophisticated trojans is possible. All this while maintaining the original target logic, with minimal impact on power and performance. Furthermore, from our exploration, we conclude that it is too naive to only utilize placement resources as a metric for HT vulnerability. This work highlights that the HT insertion success is a complex function of the placement, routing resources, the position of the attacked nodes, and further design-specific characteristics. As a result, our framework goes beyond just an attack, we present the most advanced analysis tool to assess the vulnerability of HT insertion into finalized layouts.

## 1 INTRODUCTION

Securing the development and manufacturing of integrated circuits (ICs) is a problem that the Hardware Security community is trying to solve [7]. As owning a foundry is not financially viable for most design houses, they have become fabless entities that have to rely on third-party foundries for manufacturing their designs. In this business model, the circuit layout is developed in-house and its manufacturing is outsourced. Exposing the layout to a third party is a potential threat to the IC's trustworthiness. A malicious individual could take ownership of this layout and manipulate it for his own purposes. Many potential threats have been discussed [30, 42], including insertion of hardware trojans (HTs), IP piracy, IC overbuilding, reverse engineering, and counterfeiting.

Even though only a few validated examples have been observed [13], the risk of a security breach due to hardware tampering has been in focus for many years [42]. Thus, in the past decade many potential vulnerabilities and possible countermeasures have been demonstrated. In this work, we focus on the feasibility of HT insertion during a fabrication-time attack.

HTs are designed to leak confidential information, to disrupt a system's specific functionality, or even to destroy the entire system [35] and have a broad taxonomy [16]. They comprise a *payload* implementing the malicious behavior and a *trigger* that ensures that the HT remains dormant until a specific condition is met. The *target* of an HT is the circuit into which the HT is inserted, e.g., a crypto-core or any other IP in a SoC design.

Several digital HT architectures have been proposed recently [42], with a few even demonstrated in silicon [13]. However, not many disclosed how their HT is inserted during the attack. Fabrication-time HT insertion in previous works relies on extensive knowledge of the victim's circuit [2, 27], for example of the security-critical nodes. In a fabrication-time attack, only the layout is available to the attacker, limiting the applicability of the approaches. The main contribution of our work is *to shed light on how successful an attack can be under the assumption of very limited information* about the target circuit.

High-level functionality reconstruction tools can be used to reconstruct the purpose of signals inside the design. For example, the finite-state machine of a target design can be recovered and control and data paths nodes can be distinguished [22]. The output of such tools can be utilized to automate the search of security-critical nodes on a given layout – this is the case in our work. Our main contribution is a full framework for inserting a HT utilizing only the target layout as information. For assessing our methodology, we utilized as targets the cryptocores AES-128, SHA-256, and RSA, and the general purpose PULPino microcontroller. For each target, multiple different HTs are selected for insertion. We report the success of insertion based on stealthiness, impact on power and performance and required time and discuss HT defense technique capabilities in context of our attack.

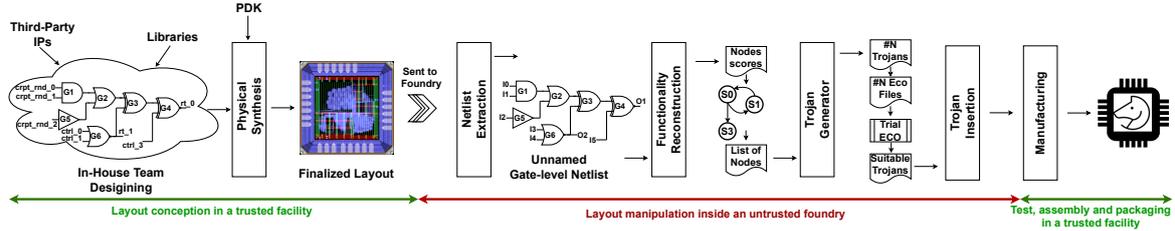

Figure 1: A typical IC design flow. Highlighted in red is the stage where a rogue element may mount an attack.

## 2 BACKGROUND
### 2.1 Related Works and Motivation

Artificial HT creation is necessary, as no public HT examples are known. The majority of published artificial HTs have been created manually (e.g., Trust-Hub benchmarks [33]). Other HTs are designed to test or overcome specific detection methods [9]. A recent survey [42] lists 27 research articles presenting HTs, but only 4 of them perform insertion of a sophisticated, non-parametric HT on the layout level. The constraints of manual HT insertion into layouts have been studied in [2]. The authors conclude that a core utilization rate of >80% will prohibit trojan insertion, but only use one HT sample with varied parameters. In [39], the authors claim that removing unused ("dead") space is sufficient to thwart HT insertion. Trippel et al. [37] analyze the susceptibility of layouts to HTs by three metrics, the unused space available for HT gates, the amount of space available for wires and the distance between HT cell placement and attacked signals. However, the authors experiment with only four HT samples and manually insert them into three fully-known IC designs. This limits the explanatory power, as we show in Section 4.

In [27], the authors propose a full framework for HT insertion during a fabrication time-attack. The authors utilized engineering change order (ECO) technique for inserting HTs, which allows the attacker to perform the attack holding only the layout database. Nevertheless, [27] makes a **strong assumption** on how the attacker searches for the security-critical nodes: the attacker can spot these nodes by visual inspection. This is not true for targets with little available information or targets with irregular placement. This limits their attack framework to a few targets, e.g., the AES core used in their demonstration.

With the rise in popularity of machine learning, various automated HT insertion frameworks have been proposed and used to provide enough HT samples to train on. Cruz et al. [4] present a HT generator for gate-level netlists. Their tool generates diverse triggers, but relies on manual insertion for payloads and does not cover the insertion at layout. In addition, trigger location is determined by a simulation-based probability metric, requiring simulation testbenches to be available. In [3] the authors improve upon the previous tool. The authors emphasize that their tool generates similar HTs to those in the training set (hence the name MIMIC). This approach might not generate a trojan set that allows generalization in machine learning, as this requires a very diverse set of trojan variants, locations, and target designs [11].

A similar tool is presented in [43], also only diversifying the trigger insertion into netlists, but improving the trigger signal selection method to be executed without a simulation testbench. Still, payload location selection is simplistic and requires manual effort.

In [32], reinforcement learning is used for HT insertion, but their targets are small ISCAS-85 benchmarks. The reinforcement learning reward driving the insertion is chosen to be dependent on the ratio of circuit inputs involved in HT activation, on the observability of the payload changes, and on the usage of at least one net with low controllability. This approach is more diverse, but tuning the reward function is a non-deterministic task requiring a high level of understanding of both the circuit and the machine learning – which is infeasible for a blind attack.

As the implementations of previous works have not been published, multiple authors of detection techniques resorted to implementing template HT generators, varying the internal structure of the HTs, but not the HT locations [11, 10].

In conclusion, existing methods lack crucial features; even automated methods require extensive knowledge about the attacked design and thus are unfit for the blind insertion attacker model. When utilizing prior knowledge, most automated tools cannot diversify payload insertion, either because payload locations must be chosen with manual effort, or because only local variation around known payload locations are possible. In this paper, we overcome the limitations by the targeted usage of reverse engineering techniques and integrate it into an end-to-end framework for blind HT insertion.

### 2.2 Threat model and Attacker Capabilities

In our framework, we assume that an attacker within the foundry has the objective of inserting malicious logic in a finalized layout. Thus, since the attacker is familiar with the manufacturing process of the foundry, he/she enjoys access to all technology and cell libraries utilized by the victim when creating the layout. We assume the attacker has no detailed knowledge about the victim's design, such as timing/power constraints, clock domains, exact functionality of the input/output pins, or high-level functionality.

For performing the attack, we assume the adversary is skilled in IC design and enjoys access to modern EDA tools including their scripting languages. From the victim's side, the attacker only has access to the layout database – typically handled in the GDSII format. For manipulating the victim's layout, we also assume the attacker knows how to apply reverse engineering techniques. Specifically, for the attack proposed in this work, the adversary has access to tools for extracting the gate-level netlist [29] and partially reconstructing the high-level functionality of the target design [22]. Furthermore, we also assume the attacker has no means to make radical modifications to the layout, e.g., manipulating the clock domains and/or changing the I/O configurations.

Designing an IC is typically executed as illustrated in Fig. 1. The layout-level work is generally considered trusted, done in-house. This is done using third-party IPs and a process design kit (PDK) together with standard cells provided by the foundry for a given technology. Here, we assume this process is trusted in which no malicious alteration is made for generating the finalized layout. The attack takes place when the victim's layout is handed over to the foundry. Precisely, the adversary is a rogue element inside the foundry that can manipulate the victim's layout before the start of the manufacturing (see the red portion of Fig. 1). This resembles attack model B as given in [41]. According to Karri et al. [16], the attacker inserts digital, gate-level HTs during fabrication time. In addition, we restrict the time when the rogue element has access to the victim's database to a full day (24 hours) [24]. This reduces attacker's capabilities to analyze and reverse engineer the victim's design, and to perform the HT insertion.

Typically, an ECO flow is used to fix small bugs in finalized layouts. ECOs are designed for post-mask modifications utilizing pre-populated gate-array cells, and pre-mask fixes, avoiding the time-consuming re-implementation of a design. However, Perez et al. [27] demonstrated that the ECO flow is a powerful tool for inserting HTs. The layout changes are local routing perturbations, as the original placement is kept intact. For inserting the HT, the ECO re-purposes empty spaces (filled with filler and spare cells[1]) with malicious logic. Since the ECO flow is executed by an industry-grade EDA tool, potential errors from manual modifications and design rule check (DRC) violations are avoided. The penalty for HT insertion is a slightly negative impact on the overall target performance due to the extra capacitance from HT wiring (see Fig. 5 for a performance comparison before and after the HT insertion). For a complete explanation of how to utilize the ECO for inserting an HT, we direct the readers to [27].

## 3 BIOHT TOOL FRAMEWORK

Our main contribution is a framework for blind insertion of hardware trojans in finalized layouts, termed BioHT. All the steps of this framework are automated; inserting HTs requires just a push of a button. Fig. 1 illustrates the BioHT framework, which comprises five distinct steps: 1) Netlist Recovery (Section 3.1), 2) Design Analysis (Section 3.2), 3) HT Netlist Generation (Section 3.3), 4) Hooking Signal Selection (Section 3.4), 5) HT insertion and Trigger Validation (Section 3.5). Fig. 2 shows the flow in detail.

### 3.1 Netlist Recovery from Layout

The attack begins once the rogue element within the foundry receives the victim's layout. First, a gate-level netlist has to be extracted from the layout. We refer to this gate-level netlist as *unnamed*, since the original hierarchy and names of cells and nets are assumed to be absent in the layout. Only the individual functionality of cells and their connectivity are recovered from the netlist extraction. Thus, in order to perform the attack, i.e., a HT-insertion with meaningful functionality, further reverse engineering is necessary.

An HT targets a specific part of the circuit, thus, **full understanding of the complete design is not necessary**. Instead of performing a full functional recovery, the attacker has to only identify security-critical signals and registers to hook the HT to. Using available information such as layout markings, datasheets, marketing material or patents, I/O-port descriptions can be inferred [28]. Global signals such as clock and reset can be identified from their connections to flip-flop pins. The resulting netlist is converted into a verilog gate-level netlist and input to design analysis. The process of netlist recovery can be seen in the top-left section of Fig. 2.

### 3.2 Design Analysis

During design analysis, *metrics* are generated to aid an adversary in searching for signals to be used for trigger and payload. This search requires a certain degree of understanding of the victim's design. For calculating those metrics, reverse engineering techniques are applied. As reverse engineering takes some time, it is a tradeoff to choose the desired level of design understanding. The metrics must be chosen carefully to keep the total runtime for the attack low.

Suitable metrics are transition probability, imprecise information flow tracking of selected signals, and the RELIC score [20]. Design analysis can be paralellized, as shown in the bottom-left portion of Fig. 2. In the remainder of this section, the calculation of these metrics is outlined.

**Transition Probability:** The HT trigger must seldomly activate in order to avoid detection during functional tests [41]. Thus, a metric is necessary to identify signals with low probability to activate the HT. In previous literature, the transition probability was introduced [31, 43]. If the transition probability is low, the probability that the required transitions for HT activation occur is also low. The transition probability is a function of the signal probability. The signal probability ($P_s(x)$) of $x$ is the fraction of clock cycles during which $x$ is 1. A transition occurs if a 1 follows a 0 or vice versa, i.e. the transition probability can be estimated as $P_t(x) = 2 \cdot P_s(x) \cdot (1 - P_s(x))$, assuming temporal independence [26].

To initialize the calculation, set $P_s = 0.5$ for the primary inputs and all flip-flop outputs. This assumption is reasonable for crypto-cores and approximately correct for other large circuits. For each combinatorial gate in the fan-out of the initialization, $P_s$ of the outputs is calculated. The output $P_s(o)$ of a 2-input or-gate is $P_s(a) + P_s(b) - P_s(a) \cdot P_s(b)$, the output $P_s(o)$ of a 2-input and-gate is $P_s(a) \cdot P_s(b)$ and the output $P_s(o)$ of a not-gate is $1 - P_s(a)$. The necessary formulas for other n-input 1-output gates can be produced by decomposing them into 2-input gates. In order to improve the probability results for flip-flop outputs, further iterations of the algorithm can be performed. In each iteration, the flip-flop input probabilities override the initial $P_s = 0.5$. Finally, the $P_t(x)$ is calculated for all signals and saved as the metric value.

**Spatial Clustering:** Co-location of HT hooking signals is desired for short wire lengths, but repeated distance calculation has a high overhead. Thus, BioHT precalculates using complete linkage clustering [25] with a $L_1$ distance metric: The target's cells are agglomerated into larger clusters. The agglomeration stops once a distance limit is reached, i.e., in the final clusters the cells remain closer than the limit. The mapping of cells to these clusters is saved as a metric.

---

[1]Typically, the placement density utilization (i.e, area with active cells versus unused area) is in the range of 50 to 60% for modern SoCs in the FinFET era. Thus, nearly half of the area of a commercial IC today can be populated by spare/filler cells.

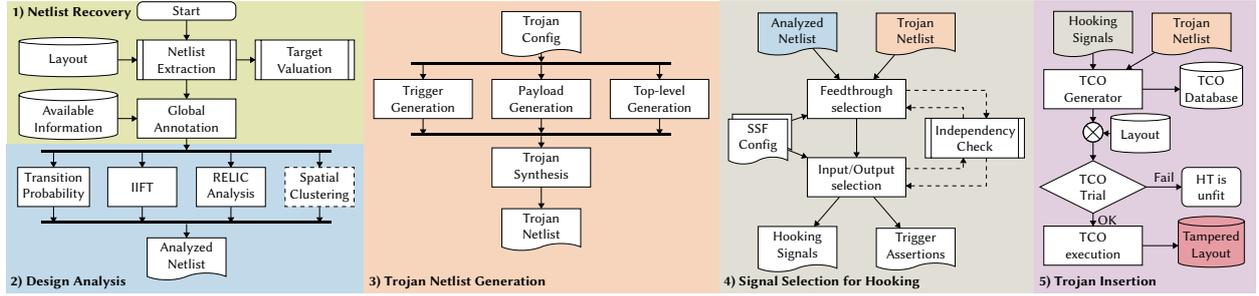

Figure 2: Steps 1)–5) of the BioHT Tool Framework explained in detail. Colored nodes represent an end-product of a previous step in the flow. The flow starts at the top left, while the tampered layout (highlighted in red) is the end-result.

**Imprecise Information Flow Tracking:** A HT that leaks information requires a metric that explains where valuable (tainted) input data can be found inside the circuit. Imprecise Information Flow Tracking (IIFT) [14] estimates an upper bound of information flow, as it assumes that each logic gate carries tainted information from its inputs to the output. The IIFT metric therefore marks as tainted any signal that exists in the output cone of the initially tainted signals. The mark, i.e., tainted or untainted, is saved as the metric value. This metric overestimates the availability of secret information at a specific signal. It must be complemented with another metric that explains the functionality of signals, such as the RELIC score.

**RELIC Scoring and FSM identification:** A valuable high-level information is the identification of whether a register (flip-flop or latch) belongs to the control logic or the data path. Using this information, HT payloads can be targeted to specific parts of the design functionality, for example to modify the control FSM or leak valuable processed data.

Finding the registers belonging to either the control or datapath logic can be performed with the NETA toolset [21]. RELIC 2 assigns every register a z-score that explains the level of dis-similarity to other registers. The underlying assumption is that data registers in a data-word show similar fan-in structure, while control registers are more unique. REDPEN returns pairs of dependent registers, i.e., a register dependency graph, in which every edge represents a path from the first register to the second register. TJSCC analyzes the register-dependency graph for its strongly connected components (SCC). As a result, there is a SCC for each register in the design. Combining RELIC and TJSCC, the SCC containing the register with the highest RELIC z-score implements the (most important) FSM in the design. Two metrics are saved for each register, its z-score and the SCC number it belongs to.

### 3.3 Hardware Trojan Netlist Generation

The BioHT HT Netlist Generator receives a configuration file, in which the user can choose any trigger/payload combinations and parameter values from our templates (see Fig. 3). Our selection of trigger and payloads covers known architectures [33, 18, 1], as well as novel payloads (i.e., leakage through FSK/DBPSK, fault sweeping). A configuration file generator is provided to ease the process of generating multiple configuration files. Nevertheless, the HT generation can be skipped, our framework is flexible enough for the user to implement their own trigger and/or payload architectures since the HT insertion only cares about the interface (i.e., where the HT is hooked to the target circuit).

For our HT architectures, we use three kinds of triggers; combinatorial, counter, and FSM; and four kinds of payloads; leak, shift'n'burn, modify, and fault. The explanation of the triggers and payloads is presented in Fig. 3. The HT triggers either wait for a *Combinatorial* condition $v$ from $n$ bits, waiting for $v$ changes of $n$ bits using a *Counter*, or traverse a $s$-state FSM on $n$-bit conditions $v_i$, masking some of the $n$ bits with $m_i$. The payloads *Leak* $n$ bits through a side channel code $M$ at a rate of $1/2^c$ bits per clock cycle, *Shift'n'burn* energy with $n$ transitions per clock cycle, *Modify* $n$-bits to a value $v$, or try to *Fault* $n$ bits by flipping them in any combination (inspired by [2]). In order to differentiate the interfacing connections, we distinguish HT ports as: input-only, output-only, and feedthrough. HT input-only ports are used for triggering or for payloads that do not drive any node. HT output-only ports drive nodes, however, do not have any corresponding input hooked to the target circuit. HT feedthrough ports are a special case, they are pairs of input-output ports that disrupt original connections between two nodes of the original circuit. Thus, the HT has control of the bit value of the disrupted connection. This type of port is utilized by the Modify and Fault payloads (see Fig. 3). The entire process of the HT Netlist Generator is shown in step 3) of Fig. 2.

### 3.4 Signal Selection for Hooking

The fourth step of the BioHT framework is to select appropriate security-critical signals to hook the HT generated in the previous step. For searching those signals, the tool requires the metrics calculated during the design analysis, and the HT interface characteristics (i.e., number of input-only, output-only, and feedthrough ports). The process starts by associating a signal selection function (SFF) for each interface port of the HT. Then, the SSF iteratively yields candidate signals from the target circuit to be hooked to each HT port, all this based on one or multiple provided metrics. The $T$ SSF yields all signals in the circuit in increasing order of transition probability, while ensuring that duplicate nets and buffer trees are avoided. This SSF is used if the total trigger probability is a sum of individual probabilities (e.g. for the *Counter* trigger). In contrast, the $TR$ SSF yields signals with low transition probability randomly using a negative-exponential weighting function. This overcomes detection tools searching for rare signals only, while providing better diversity than a threshold-based approach [4]. These SFFs might require the additional use of the spatial clustering metric (TC and TCR) if signals selected by T and TR are spatially distant. Further SSFs are tailored to selecting payload signals. An explanation of SSFs implemented, and their behavior is given in Tab. 1.

Selection of signals to hook the HT to has to be meticulously done, and it is not only a function of the calculated metrics. It is highly desirable to avoid mutually dependent signals: If a signal used for triggering is dependent of an active payload signal, a combinational loop could be generated. Also, hooked signals used as Modify or Fault payloads should be independent, for maximizing the effectiveness of the HT.

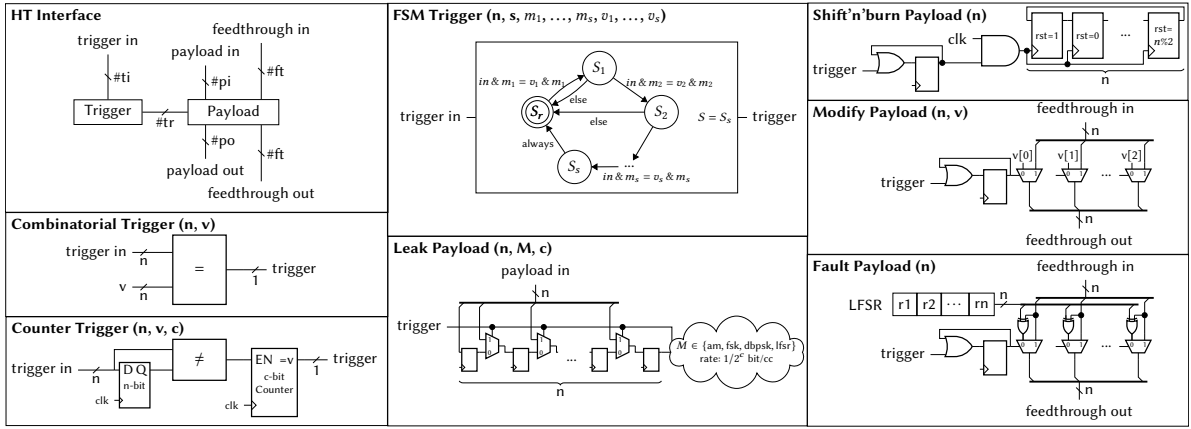

Figure 3: HT Interface and available trojan triggers and payloads. Trigger and payload parameters are given in parentheses.

Table 1: Available Signal Selection functions

| Function | metrics used | Behavior |
| --- | --- | --- |
| T | Trans. Prob. | sample low probability signals |
| TC | Trans. Prob., Spat. Clust. | sample low probability signals from adjacent clusters. |
| TR | Trans. Prob. | randomly sample low probability signals. |
| TCR | Trans. Prob., Spat. Clust. | randomly sample low probability signals from adjacent clusters. |
| RLR | RELIC z-score | randomly sample low z-score (i.e. data) signals |
| RLT | RELIC z-score, IIFT | sample tainted signals with z-score below threshold |
| RHS | RELIC z-score, SCC | sample signals from the highest-z-score SCC (i.e. FSM) |
| RHST | RELIC z-score, SCC, IIFT | sample signals from the tainted FSM (e.g. tainting instr. memory) |
| D | — | Connect no signal to this I/O. (e.g. shift'n'burn feedback signal) |

For avoiding the situations described above, our signal search engine repeatedly checks the dependencies of each candidate signal. Only independent signals are considered for hooking the HT. To check the independence of two signals, a directed acyclic graph representation of the target design is created by replacing all registers by virtual input and output ports. Two signals are deemed independent if they do not have a common ancestor in the directed acyclic graph representation. This approach increases the amount of available trigger input signals significantly compared to the topological order approach used in other works [4, 43].

### 3.5 Trojan Insertion

For inserting the HTs into the victim's layout, we utilize an ECO flow similar to [27], as described in Section 2.2. In our framework (see Fig. 1), we approach the ECO differently than the one demonstrated in [27]. Instead of directly modifying the victim's netlist, we utilized the ECO file format. This format is supported by EDA tools and procedurally describes the modifications to be performed by the ECO. The advantage of utilizing this format is doing the ECO interactively. It is necessary to load the design only once for analyzing multiple ECO files before committing the modifications. By doing this, the adversary significantly saves execution time, since loading large designs can take several hours. Instead, loading an ECO file takes a few minutes. Before committing, the adversary can check if the HT fits in terms of placement resources and timing with a *trial insertion*.

To make the attack even faster, we introduce the concept of Trojan Change Order (TCO) format file. The TCO file is generated from the signals to hook the HT in combination with the HT netlist. We utilize the same syntax as the ECO file, however, with commented lines containing directives for the BioHT tool. Those directives are used to configure the type of HT (e.g., leak, deplete, modify or fault), number of connections and location of the HT gate-level netlist. Thus, it is possible to pre-generate TCO files for several types of HT, and specialize it according to the target's evaluation. Thus, it is feasible to create a database of HTs rapidly available for an attack, as shown in Fig. 2. Finally, if the HT fits, the attacker can commit the changes with a final TCO execution.

Trigger validation is performed after HT insertion by adding the respective SystemVerilog assertions (Trigger Assertions in Fig. 2) to the final netlist module and using any formal verification tool capable of handling cover property assertions. Besides the netlist, modern formal verification tools require only little additional information to prove HT triggering. Clock and reset inputs and polarities are found automatically, as well as most initialization sequences. If available from the design analysis step, the user can specify additional formal verification constraints and constants, e.g., for scan enable pins. If the validation is successful, the formal verification tool provides the attacker with a testbench to trigger the HT.

## 4 BIOHT FRAMEWORK EVALUATION AND RESULTS

In this section, we demonstrate our methodology and the settings utilized for evaluating the BioHT framework. All experiments performed in this work use industry-grade EDA tools. The layouts generated during the experiments are tapeout-ready, meaning they have proper power planning, timing closure, and no design rule violations.

**Targets, Evaluation Settings and Procedure:** For evaluating the BioHT capabilities, we have utilized four designs as targets – three crypto cores and one microcontroller. We chose the crypto cores AES-128, SHA-256, and RSA. The microcontroller utilized is the open-source SoC PULPino featuring one 32-bit RISC-V core [36]. The PULPino uses multiple memories and shows that BioHT can insert into large and complex targets.

For the implementation, we have utilized a commercial 65nm CMOS technology with a nominal voltage of 1.2V. For timing the designs, we utilized the slow process corner (SS), temperature of 125°C, and under voltage of 1.12. This is the worst case setup corner recommended by the vendor. For reporting the power consumption, we utilized the typical process corner (TT), temperature of 25°C, and the nominal voltage of 1.2V. These practices are in line with the standard flow adopted by the IC industry.

The first step is to generate a tapeout-ready layout for each target as the victim would do. For PULPino, RSA, and AES-128, we balanced density and performance, with the goal to achieve high-density designs operating at a considerably fast clock. For SHA-256, we set the density to a less ambitious value (i.e., around 50%) and then aimed for maximum performance. The results are presented in Tab. 2. It was possible to achieve both high-density and high-speed for AES-128 and RSA, above 80% of utilization and a clock frequency of 750MHz. In contrast, PULPino's density and clock frequency are low due to the presence of 8 memories (see Fig. 4). However, the achieved frequency of 285MHz is competitive with other PULPino implementations in a similar 65nm CMOS technology [6].

We extracted the gate-level netlist from the layout, estimated the operating frequency of the target and the power consumption (see Tab. 2), similarly as done in [27]. Steps 2)–4) were implemented in Python using several open-source libraries [8, 34, 23, 5]. Note that for all targets except PULPino, we assumed that signal tainting was only possible for the I/O. In PULPino, for the HT variant with Combinatorial trigger and Modify payload, we assumed that the attacker has acquired additional information, for example from an insider. This shows that our framework is capable to adapt to any additional available data about the design under attack. We assume that the data allows to taint the core control FSM registers, so that the payload can insert random pipeline flushes for performance degradation. After generating the TCO files, the actual HT-insertion is performed. To demonstrate the boundaries of HT insertion, we also executed unfit HTs that lead to trivially tampered layouts. Evaluation of the performance of the tampered layouts was done **using signoff settings** and provides accurate performance impact.

**Experimental Results:** To show the diversity of achievable HTs using BioHT, we performed the insertion of 96 variations of HTs. In the remaining results presentation, we are going to show only a handful of the possible HTs, the full results for all HT variants can be found at [12]. The results in Tab. 2 show the injected type of trigger and payload, the number of sequential/combinational cells, and the number of connections that the HT hooks to the target circuit. As mentioned before, BioHT can generate a vast number of HTs when employing all configuration options and trigger-payload combinations. Depending on the available target information, the attacker can narrow the range of fitting configurations and trigger-payload combinations. For example, a bit-flip fault as produced by the fault payload fits a crypto-core better than PULPino. Parametrization was guided by Trial TCO, as well as the independency check. For example, the independency check highlighted for both AES and RSA that there is no FSM, as only one independent FSM state register was found. This is why the Modify payload was not used for FSM tampering with these targets. Trial TCO guided towards low register count (that is lower $n$ and $c$ parameters), as register count was directly related to unfittability. Note that, among the selected HTs presented in this work are smaller and larger possible HTs. The SSFs were selected so that their behavior (see Tab. 1) fits the HT. We can conclude that BioHT is capable to produce diverse HTs for virtually any given target, while adapting to the amount of available information.

One goal of HT insertion is to guarantee that layout performance is not affected. The impact on the target layout is shown in Tab. 2 in terms of density, total power, timing (critical path slack), and design rules violations. An actual attacker would drop the HT variants that show unfit timing (large negative slack) or design rules violations (e.g. the last RSA variant). For all other variants, the impact of their insertion is not enough for breaking the design, making the resulting tampered layout manufacturable.

We selected the targets PULPino and RSA for demonstrating the HT insertion results in more detail. The HT variant with a Counter trigger and a Modify payload is challenging for its large number of cells (highlighted in bold in Tab. 2). The contrast between the layouts before and after the HT insertion is depicted in Fig. 4, where the malicious cells are highlighted in red, and the hooked cells are highlighted in light blue. For this image, we omitted the filler cells to improve the contrast between the original cells and HT cells.

The impact on the timing performance is demonstrated in Fig. 5. The green bars show the distribution of the timing slack of paths before and the red bars after inserting the HT. As portrayed in this figure, the overall impact is greater on the denser design, i.e., RSA. This is expected since the increase in the routing congestion is lesser in low-density designs. This claim is supported by the wire length statistics reported in Tab. 3[2]. From these results, it is clear that routing the additional HT logic is easier in PULPino (low-density design), because the additional wiring can be evenly distributed in all metal layers. The opposite is true for RSA (high-density design), where it was required to use an additional layer that was empty before, and in M5 more than the double amount of routing metal was used. Therefore, the timing impact on RSA from the additional coupling capacitance is stronger.

For this work, we executed further fit and unfit HT TCOs for demonstrating the dependency between the overall design implementation, HT architecture, and, number/location of chosen signals to hook the HT. Tab. 2 shows that the density of the design is not the only factor that makes an HT unfit. In the case of the AES-128, even with a density of 84.54%, all HTs were inserted without breaking the target design. This contradicts previous works that stated it would be impossible to insert HTs in designs with 80% of density or higher [2], or being a very difficult task if the placement resource available is small [37]. The size of the HT and the number of hooked signals are also not a sufficient metric to decide if an HT is unfit. Still using AES-128 as an example, the total timing impact does not depend on those HT characteristics, where the large timing impact is not from the largest HT. Therefore, it is not possible to rely on these metrics to assess with precision the vulnerability of a layout against HT insertion. Another experiment on PULPino was performed using the cluster-based SSFs (TC, TCR). By enforcing co-location of hooks, all PULPino variants become feasible. We conclude that the feasibility to co-locate HT signals (for a given attack objective) has a large impact on layout vulnerability.

---

[2]In the 9-metal stack used in our experiments, M1 cannot be used for signal routing. For this reason, M1 is not shown. Similarly, M8/M9 are reserved for power distribution and are not shown.

When developing BioHT, an important aspect was to ensure a runtime within the 24h time window available to perform the attack. In Fig. 6, we show the runtime required for our attack in contrast with the design implementation runtime for PULPino and RSA. For PULPino, we show the HT-insertion with and without the use of the Spatial Clustering (optimized) Metric for searching the hooking signals. Trial TCO can easily determine if the overhead for spatial clustering is required. The runtime for RSA, PULPino with optimization, and PULPino without, are 50 minutes, 44 minutes, and 24 minutes, respectively. As shown in the Fig. 6, the most time intensive tasks are the netlist extraction and power/time analysis. Those tasks are mandatory if the intention of the attacker is to insert meaningful and stealthy HTs. Our tool BioHT represents less than 20% of the attack total runtime. We conclude that HT generation with BioHT achieves competitive runtime overhead and would allow repeated attack execution for optimum performance.

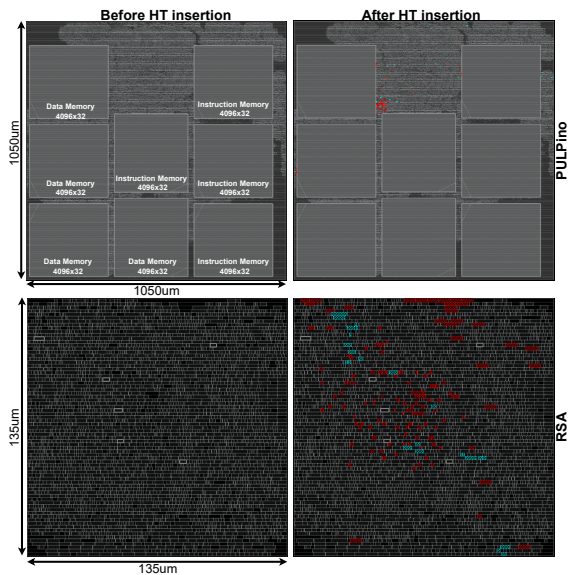

Figure 4: Layout contrast before and after HT insertion for PULPino and RSA. The HT inserted has a trigger counter and a modify payload. Its cells are highlighted in red.

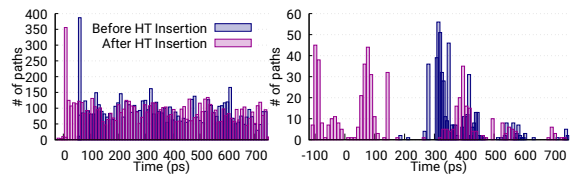

Figure 5: Timing impact of the hardware trojan Insertion, for PULPino (left) and RSA (right).

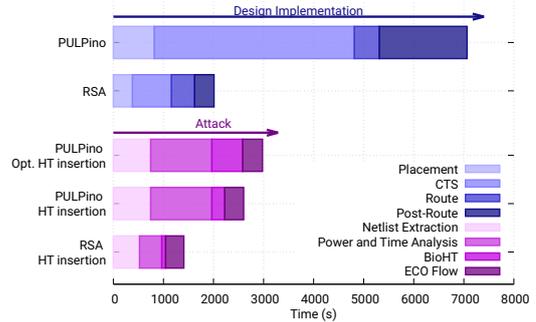

Figure 6: HT insertion using BioHT execution time (s) in contrast with the physical implementation of a target.

## 5 DISCUSSION

During the experiments, the ease of the BioHT flow proved valuable, as few configuration options have to be adapted to run HT insertion on a new design. Configuration was also guided by the tool itself. For example, the independency check proved to have a significant advantage over mere topological ordering. It aids the exploration of possible HT configurations, as it immediately highlights that specific configuration parameters are infeasible. It must be noted, however, that the experiments showed that, in rare cases, the independency check cannot guarantee that a HT can be triggered. In this case, the user is warned and recommended to use a different random seed for the trigger SSFs configuration, or select a different SSF. In our experiments, this occured once and was solved with a new random seed immediately.

An important and unforeseen result is that the influence of circuit cell density on the feasibility of HT insertion is much less than expected. In the two low density designs, SHA and PULPino, the HT insertion partially failed, while in the high density designs HT insertion succeeded, even for large HTs with hundreds of cells, and independent of the increase in wire length. We conclude that the vulnerability of a layout for HT insertion cannot be assessed by a few metrics. Instead, we propose to use BioHT as an empirical solution to assess the vulnerability by random HT insertion. BioHT goes beyond a proof-of-concept that blindly attacking a layout is possible. The framework can quickly produce a boundary of HT insertion feasibility, provide a risk assessment and guide physical defense strategies for HT insertion. Our claim is that no other work in the literature can provide this information.

In order to defend against sophisticated foundry-level attacks as in this work, two possibilities are pre-silicon design-for-trust or post-silicon detection techniques. Pre-silicon design-for-trust implements measures to render HT insertion more difficult or to provide trust anchors in the design to identify tampered locations. For example, logic locking renders understanding an unknown layout more difficult [17]. As previously explained, little available information is enough to insert meaningful HTs with BioHT. In how far this limits the effect of locking remains as a future work. Other design-for-trust techniques such as on-chip sensors must be circumvented, but automated solutions exist for many techniques (e.g. [40]), which are easily integrated into BioHT.

Post-silicon detection generally is either an exhaustive, error-prone and costly reverse engineering effort [19] to fully inspect the design, or can by principle only achieve partial coverage (e.g., with side-channel analysis, logic testing, etc.) [17]. BioHT can be adapted to evade non-exhaustive post-silicon detection by modifying HT configuration and SSFs. An interesting approach is to include self-test structures in pre-silicon to improve post-silicon imaging detection [38]. In the end, post-silicon detection tries to prove that the design is identical to the pre-silicon layout, but a solution with guaranteed coverage is yet to be found.

Table 2: Physical synthesis results before and after HT insertion for the target designs and several HT architectures.

| Target | Trigger | Payload | SSF Trig. | SSF Payl. | Seq. # | Comb. # | Conn. # | Freq. (MHz) | Density (%) before | Density (%) after | Total power (μW) before | Total power (μW) after | Slack (ps) before | Slack (ps) after | Viol. # after |
|---|---|---|---|---|---|---|---|---|---|---|---|---|---|---|---|
| PULPino | FSM | Modify | TR | RHS | 7 | 14 | 20 | | | 54.00 | | 72.20 | | -4 | 0 |
| | FSM | Shift'n'burn | TR | D | 40 | 75 | 6 | | | 54.23 | | 72.59 | | -8 | 0 |
| | Comb. | Modify | TR | RHST | 4 | 8 | 19 | | | 53.99 | | 72.48 | | -9 | 0 |
| | Comb. | Modify | TCR | RHST | 4 | 15 | 31 | | | 54.03 | | 72.77 | | 2 | 0 |
| | Comb. | Shift'n'burn | TR | D | 37 | 71 | 8 | 285 | 53.97 | 54.26 | 74.64 | 72.61 | 15 | -10 | 0 |
| | **Counter** | **Modify** | **T** | **RHS** | **24** | **79** | **18** | | | **54.33** | | **72.52** | | **-18** | **0** |
| | Counter | Modify | TC | RHS | 24 | 82 | 15 | | | 54.43 | | 73.00 | | 0 | 0 |
| | Counter | Shift'n'burn | T | D | 57 | 140 | 3 | | | 54.51 | | 72.99 | | -46 | 0 |
| | Counter | Shift'n'burn | TC | D | 57 | 140 | 3 | | | 54.66 | | 73.04 | | 3 | 0 |
| SHA-256 | FSM | Modify | TR | RHS | 24 | 85 | 28 | | | 54.38 | | 47.27 | | 20 | 0 |
| | FSM | Shift'n'burn | TR | D | 40 | 75 | 6 | | | 54.54 | | 47.28 | | 31 | 0 |
| | Comb. | Modify | TR | RHS | 4 | 18 | 38 | | | 53.94 | | 47.28 | | -7 | 0 |
| | Comb. | Shift'n'burn | TR | D | 37 | 71 | 8 | 500 | 53.84 | 54.58 | 46.48 | 51.74 | 49 | 34 | 0 |
| | Counter | Modify | T | RHS | 24 | 85 | 25 | | | 54.38 | | 47.27 | | 20 | 0 |
| | Counter | Shift'n'burn | T | D | 57 | 140 | 3 | | | 55.03 | | 51.94 | | 45 | 0 |
| RSA | FSM | Fault | TR | RLR | 13 | 31 | 30 | | | 87.40 | | 15.58 | | -84 | 0 |
| | FSM | Shift'n'burn | TR | D | 40 | 75 | 6 | | | 89.62 | | 15.73 | | 139 | 0 |
| | Comb. | Fault | TR | RLR | 10 | 29 | 33 | | | 87.21 | | 15.87 | | -46 | 0 |
| | Comb. | Shift'n'burn | TR | D | 37 | 71 | 8 | 750 | 86.19 | 89.39 | 15.58 | 15.62 | 196 | 194 | 0 |
| | **Counter** | **Fault** | **T** | **RLR** | **30** | **96** | **24** | | | **89.34** | | **15.90** | | **-108** | **0** |
| | Counter | Shift'n'burn | T | D | 44 | 109 | 3 | | | 91.57 | | 15.97 | | N/A | 71 |
| AES-128 | FSM | Fault | TR | RLR | 13 | 31 | 28 | | | 84.85 | | 21.96 | | 76 | 0 |
| | FSM | Shift'n'burn | TR | D | 40 | 75 | 6 | | | 85.85 | | 22.11 | | 30 | 0 |
| | Comb. | Fault | TR | RLR | 10 | 29 | 38 | | | 84.79 | | 21.95 | | 52 | 0 |
| | Comb. | Shift'n'burn | TR | D | 37 | 71 | 8 | 750 | 84.46 | 85.55 | 21.82 | 22.12 | 201 | 116 | 0 |
| | Counter | Fault | T | RLR | 30 | 96 | 24 | | | 85.99 | | 22.15 | | 119 | 0 |
| | Counter | Shift'n'burn | T | D | 10 | 29 | 2 | | | 84.78 | | 21.94 | | 163 | 0 |

Table 3: Routing length in μm per metal layer for PULPino and RSA, before and after the HT-insertion.

| Metal layer | RSA before | RSA after | PULPino before | PULPino after |
|---|---|---|---|---|
| M2 | 18.7k | 18.9k | 323.8k | 326.5k |
| M3 | 23.2k | 25.1k | 439.2k | 441.7k |
| M4 | 11.5k | 15.9k | 378.2k | 382.4k |
| M5 | 2.5k | 5.1k | 357.1k | 360.6k |
| M6 | - | 1.0k | 221.3k | 223.8k |
| M7 | - | - | 161.6k | 163.2k |

As BioHT resembles a real-world blind foundry-level attack, the framework can complement post-silicon inspection by guiding inspection to vulnerable locations. This reduces the effort required for inspection, as only selected regions of the design must be inspected, namely those where BioHT potentially attacked the circuit.

Orthogonally, the use of machine-learning in any pre- and post-silicon HT detection is on the rise [17, 15]. Providing a vast amount of diverse HT samples is essential to train any machine-learning based detection technique. BioHT can be used blindly on any design to generate diverse training samples. Unlike with previous methods, it is not necessary to manually implement a sample set of trojans or to perform manual adaptation of the HT insertion technique. We expect the diversity of HTs and targets to be necessary and beneficial to future work on machine-learning based HT detection.

## 6 CONCLUSION

In this work, we presented BioHT, a framework to insert hardware trojans into unknown ASIC layouts. The tool is using an end-to-end approach, starting at the victim's layout delivered to the foundry and ending with the tampered layout ready for fabrication. Through the use of reverse engineering techniques, little knowledge about the design is necessary to introduce sophisticated trojans into the circuit. The use of state-of-the-art physical synthesis tools and an ECO flow allows performing the actual insertion as trouble-free as a layout bug fix. The end result is a DRC-clean layout that is ready for manufacturing, or an error message to the attacker that the parameters of insertion must be changed. Our experiments show that the complete approach can be executed multiple times in the time-frame between tape-out and manufacturing, so that the optimum trojan could be selected out of several possibilities. However, BioHT is also a tool to show how realistic trojan insertion would be performed, and can guide risk assessment, defense, and future research. To conclude, HT insertion in finalized layouts is a real threat to today's globalized IC manufacturing and must not be taken lightly. Our framework provides the necessary capabilities and information to advance countermeasures against this threat.

## ACKNOWLEDGMENTS

This work has been partially conducted in the project "ICT programme" which was supported by the European Union through the European Social Fund. It was also partially supported by European Union's Horizon 2020 research and innovation programme under grant agreement No 952252 (SAFEST).